\documentclass[10pt,twocolumn]{article} 
\usepackage{simpleConference}
\usepackage{times}
\usepackage{subcaption}
\usepackage{graphicx}
\usepackage{amssymb}
\usepackage{orcidlink}
\usepackage{authblk}
\usepackage{biblatex}
\usepackage{url,hyperref}

\providecommand{\keywords}[1]
{
  \small	
  \textbf{\textit{Keywords---}} #1
}

\usepackage{titlesec}
\titleformat*{\subsection}{%
    \fontsize{10}{12}\bfseries%
}

\begin{document}

\title{SHREC'25 Track on Multiple Relief Patterns: Report and Analysis}

\makeatletter
\renewcommand*{\@fnsymbol}[1]{\ensuremath{*}}
\makeatother
\newcommand*\samethanks[1][\value{footnote}]{\footnotemark[#1]}

\author[1]{Gabriele Paolini\thanks{\raggedright Track organizers. Corresponding author: \texttt{gabriele.paolini1@unifi.it}}\orcidlink{0009-0003-5325-2076}}
\author[2]{Claudio Tortorici\samethanks\orcidlink{0000-0001-6943-2854}}
\author[1]{Stefano Berretti\samethanks\orcidlink{0000-0003-1219-4386}}
\author[3]{Ahmed Hazem Youssef}
\author[3]{Halim Benhabiles\orcidlink{0000-0003-2950-5312}}
\author[4]{Adnane Cabani\orcidlink{0000-0001-5948-8950}}
\author[5]{Ruiwen He\orcidlink{0000-0002-5381-6068}}
\author[6]{Karim Hammoudi\orcidlink{0000-0002-4804-4796}}
\author[7]{Iyyakutti Iyappan Ganapathi\orcidlink{0000-0001-6312-5765}}
\author[7]{Syed Sadaf Ali\orcidlink{0000-0002-0198-8319}}
\author[7]{Divya Velayudhan\orcidlink{0000-0003-2897-570X}} 
\author[7]{Maregu Assefa\orcidlink{0000-0003-2815-7993}} 
\author[7]{Naoufel Werghi\orcidlink{0000-0002-5542-448X}}

\affil[1]{Media Integration and Communication Center, University of Florence, Florence, 50134, Italy}
\affil[2]{Technology Innovation Institute, Abu Dhabi, 9639, United Arab Emirates}
\affil[3]{IMT Nord Europe, Institut Mines-Télécom, Université de Lille, Center for Digital Systems, F-59000 Lille, France}
\affil[4]{ESIGELEC, IRSEEM, Université de Rouen Normandie, 76000 Rouen, France}
\affil[5]{Léonard de Vinci Pôle Universitaire, Research Center, 92916, Paris La Défense, France}
\affil[6]{Department of Computer Science, Institut de Recherche en Informatique, Mathématiques, Automatique et Signal, Université de Haute-Alsace, 68093 Mulhouse, France}
\affil[7]{Khalifa University of Science and Technology, Abu Dhabi, United Arab Emirates}

\maketitle
\thispagestyle{empty}

\begin{abstract}
 This SHREC 2025 track focuses on the recognition and segmentation of relief patterns embedded on the surface of a set of synthetically generated triangle meshes.
We report the methods proposed by the participants, whose performance highlights the inherent complexity of solving the problem, which is still open.
Then, we discuss the critical aspects of the proposed tasks, highlight the limitations of current techniques, and outline possible directions for future research.
All resources and track details are available at the \href{https://sites.google.com/unifi.it/shrec25-relief-pattern}{\color{blue}official track webpage}.
\end{abstract}

\keywords{Relief Pattern, 3D Segmentation, 3D Retrieval}

\section{Introduction}
This Shape Retrieval Challenge (SHREC 2025) track on ``Retrieval and Segmentation of Multiple Relief Patterns'' was designed to evaluate the performance and promote the development of novel methods for the segmentation and retrieval of relief patterns over surfaces.
Unlike previous contests primarily focused on retrieval tasks~\cite{SHREC17, SHREC18, MOSCOSOTHOMPSON2020199}, this track simultaneously assesses both the segmentation and retrieval capabilities of the proposed automatic algorithms, thus offering a more comprehensive benchmark for this class of problems.
Participants are required to first identify regions corresponding to distinct relief patterns and subsequently retrieve, from a target dataset, all surfaces exhibiting similar patterns.

A relief pattern (also referred to as a 3D texture or geometric texture) consists of a local and repeated deformation lying on a 3D surface~\cite{giachetti2018effective}.
Relief patterns are characterized by being independent from the global structure of the object on which they are defined.
Typical examples include decorative engravings on pottery, artifacts, and monuments, textures on fabrics, and tree bark surfaces~\cite{9611230, BLUSSEAU2022110333,OTHMANI20132144}.
Relief patterns can convey crucial information regarding the object's material properties, its function, and cultural significance.
In order to properly analyze them, it is necessary to focus on the local geometric features of a surface rather than on its overall shape.
This aspect, combined with the limited availability of (geometric) textured 3D datasets, has contributed to the relatively scarce scientific exploration of this class of problems~\cite{Ganapathi2022LabeledFN}.

Compared to previously published tracks on relief pattern analysis~\cite{SHREC17, SHREC18, MOSCOSOTHOMPSON2020199}, this track is distinguished by three main characteristics:
\emph{(i)} the surfaces in the dataset may contain one, two, or three geometric textures, which can cover either a part or the entire surface;
\emph{(ii)} both regular and quasi-regular relief patterns are included, representing features of organic surfaces and accounting for imperfections;
\emph{(iii)} the full dataset consists of approximately 1,000 triangular meshes exhibiting complex topologies and self-occlusions.
Additionally, the proposed dataset contains disjoint sets of patterns, which is crucial for evaluating the generalization capability of the algorithms to unseen geometric textures.

The track initially recorded five registrations, confirming the relevance of the proposed problem.
Despite this initial interest, only one group submitted a partial solution addressing the original pattern retrieval problem.
A second group was unable to complete their submission, but provided important feedback on the development of their method, a brief description of which is included in this report.
This outcome mirrors the experience of the SHREC 2018 track ``Recognition of geometric patterns over 3D models'', and highlights the intrinsic difficulty of the tasks proposed~\cite{SHREC18}.
At present, no existing algorithms are capable of effectively performing segmentation or retrieval under the general conditions imposed by the dataset released within this track.

\begin{figure*}[!ht]
\centering
\includegraphics[width=0.9\linewidth]{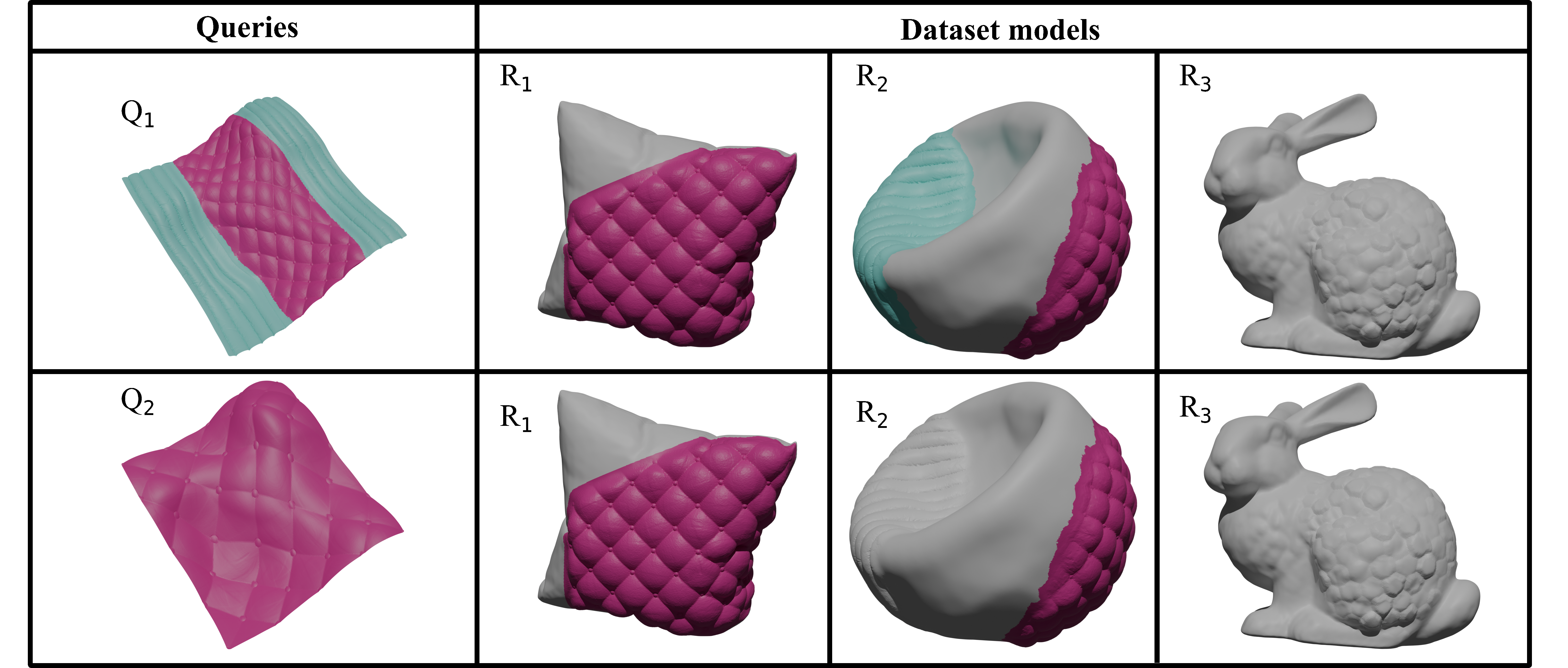}
\caption{
An example of the expected results for the retrieval task. The faces of each retrieval model R\textsubscript{i} shown in the figure are colored according to the type of pattern shared with the individual query mesh Q. Mesh R\textsubscript{1} shares a single pattern with both query meshes Q\textsubscript{1} and Q\textsubscript{2}. Mesh R\textsubscript{2} shares both patterns with Q\textsubscript{1} (pink and cyan regions), but only one with Q\textsubscript{2} (pink region only). Finally, no face of R\textsubscript{3} is annotated, as it does not share any pattern with the queries.}
\label{fig:retrieval-example}
\end{figure*}

This paper provides a detailed description of the track, including the context and motivations behind the study of relief patterns, the dataset and ground truth provided to participants, and the definition of the two tasks to be addressed.
After introducing the methods and results submitted by the participating groups, we will discuss the main challenges encountered.
The main goal of this work is to identify the open problems that remain in this field of 3D analysis, to outline the most promising research directions, and to highlight the opportunities that the published benchmark offers to the community.

\subsection{Motivation}\label{sec:motivation}
The analysis and recognition of geometric textures is an open problem that naturally evolved from earlier work on satellite imagery, aerial scans, and biomedical imaging~\cite{ijgi8100450, 9548802}.
Among the first examples of 3D datasets explicitly delivered for the recognition and transfer of geometric textures, we can mention the ``MIT CSAIL Textured Models Database''~\cite{mitcsail2008}.
Based on this dataset, Mertens et al.~\cite{Moeller-2006} proposed a method to transfer texture images between different 3D models in a visually coherent manner.
To do this, they extracted geometric features and exploited correlations in the local surface geometry to guide the transfer in a visually coherent manner. 

Over the past decade, several challenges within SHREC have addressed the retrieval and recognition of relief patterns.
Starting with SHREC 2017~\cite{SHREC17}, which focused on retrieving scanned surface patches exhibiting similar reliefs, subsequent tracks such as SHREC 2018~\cite{SHREC18} and SHREC 2020~\cite{MOSCOSOTHOMPSON2020199} placed increasing emphasis on the realism and complexity of both datasets and tasks.
In particular, the SHREC 2018 track proved challenging for multiple reasons, including the high resolution of certain models, which made per-facet analysis computationally intensive, and the realistic nature of the patterns, derived from scans of actual artifacts.
The various approaches proposed over the years highlighted the difficulty of analyzing relief patterns independently of the noise from the acquisition process, and in a manner that remains efficient even at high resolutions.

Relief patterns provide essential insights into the function, origin, and material properties of a 3D object.
In cultural heritage, the analysis of reliefs plays a fundamental role in the restoration and reassembly of fragmented artifacts and engravings~\cite{9611230,7539584}.
During archaeological excavations, pottery objects, statues, and architectural fragments are often broken or dispersed.
Sometimes these fragments are stored in different collections and locations, making digital restoration techniques essential.
In this context, local geometry can serve as a fingerprint to guide the reconstruction of fragmented objects.
Matching fragments based on the continuity or similarity of reliefs allows restorers to propose plausible assemblages in situations where the global shape is unclear.
In industrial contexts, strict control over surface properties is often required, and detecting local deviations of the surface becomes crucial for quality assurance.
An interesting application concerns the editing of reliefs (not necessarily repeated) to modify or transfer surface details onto a model before fabrication~\cite{Liu2006,Liu2007,Liu2007Background}.
In biomedical imaging, the analysis of 3D textures can provide crucial information about the structural and functional properties of biological tissues, which often exhibit spatially periodic or fractal-like characteristics~\cite{DEPEURSINGE2014176}.
Advances in texture analysis techniques directly contribute to improving diagnostic accuracy and supporting the development of personalized treatment plans.

\subsection{Problem statement}\label{sec:problem}
This track was designed to investigate two complementary but distinct aspects of relief pattern analysis: pattern retrieval and pattern segmentation.

\paragraph{Pattern retrieval} 
This task consists of identifying all 3D models that share at least one type of relief pattern with those present on the surface of a given query model.
Once the relevant models have been identified, the corresponding patterns must also be localized on their surfaces.
In the case of queries containing multiple relief patterns, any model exhibiting at least one of these patterns must be retrieved.
This process is repeated for each model in the query dataset.
Figure~\ref{fig:retrieval-example} schematically illustrates the recognition of relief patterns present on two query models and their retrieval across a set of models from the target dataset.

\begin{figure}[!ht]
\centering
\includegraphics[width=\linewidth]{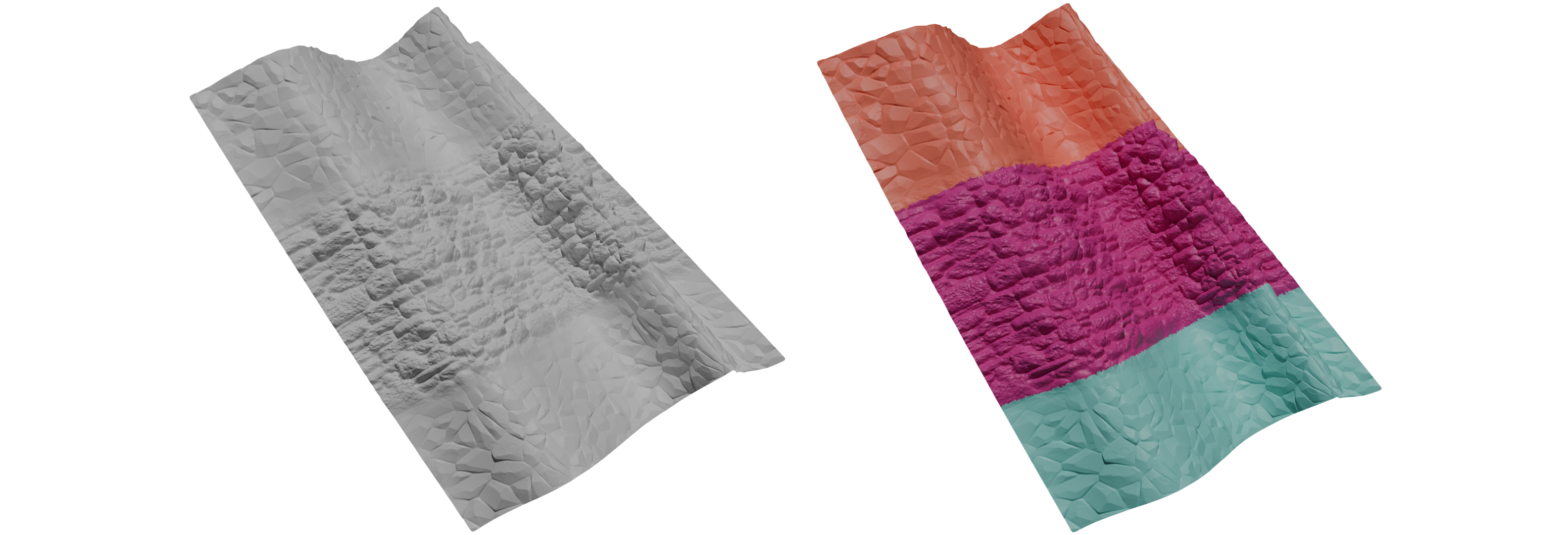}
\caption{
Example of a segmentation task result. The query mesh on the left contains two different types of relief patterns, which divide the surface into three distinct regions. On the right, the same mesh is shown with the three segmented regions highlighted in different colors.}
\label{fig:segmentation-example}
\end{figure}

\paragraph{Pattern segmentation} 
This subproblem requires participants to analyze surfaces containing more than one type of relief pattern, with the goal of generating a distinct segmentation mask for each pattern.
Segmentation is performed on the models included in the query dataset.
This task was intentionally kept optional for the evaluation of the proposed methods.
Figure~\ref{fig:segmentation-example} shows an example of segmentation results in a model from the query set.

The relevance of each 3D model with respect to the patterns present in the queries was determined automatically during dataset generation and is assigned on a per-face basis.

The choice of a track structured into two subproblems reflects a realistic scenario, where relief patterns must first be recognized and isolated before performing subsequent operations, such as retrieval within a 3D model database based on local surface characteristics.

\subsection{Paper organization}\label{sec:report}
The remainder of the paper is organized as follows.
In Section~\ref{sec:dataset}, we describe the benchmark proposed for the contest, including the ground truth, query set, retrieval set, and training set.
Subsequently, Section~\ref{sec:results} details the methods submitted for the contest, along with the metrics used to evaluate the different experiments.
In Section~\ref{sec:analysis-challenges} and ~\ref{sec:dataset-challenge}, we discuss the challenges associated with the recognition and localization of patterns on 3D surfaces.
Many of these challenges arise from intrinsic characteristics of the dataset proposed for the track and highlight the limitations of current methods developed for surface analysis.
This consideration represents an important aspect for understanding the design choices behind the dataset generation and clarifies the objective of providing a testing ground for the development of new and innovative algorithms.
In Section~\ref{sec:directions}, we explore several potential research directions that we consider relevant to the development of effective pattern analysis techniques.
Finally, Section~\ref{sec:conclusions} concludes the report with final remarks.

\section{Dataset and ground truth}\label{sec:dataset}
All 3D models are triangle meshes resulting from the automatic application of geometric textures onto synthetic base models (i.e., models not derived from 3D scans).
The choice of using synthetic models instead of scanned surfaces with existing textures is motivated by several factors.
First, synthetic models typically have smooth surfaces free from noise and artifacts, which instead are common in objects acquired through 3D scanning.
This guarantees greater control during the editing phase over the type of relief generated, as well as its spatial extension and placement.
Second, base models can be deliberately selected according to their geometric features, such as the genus, closedness or openness, self-occlusions, and local geometric details.
All these factors can be taken into account during the evaluation of automatic segmentation or retrieval algorithms.
Finally, one of the main difficulties in creating benchmarks lies in the definition of ground truth.
For 3D models obtained from scans, such as archaeological artifacts, annotation must be supported by experts and can be a labor-intensive and time-consuming process.
The use of an automatic geometric texture generation algorithm has the advantage of producing ground truth information automatically and with the highest possible fidelity with respect to the relevant regions.
A direct consequence of this approach is the possibility of creating large-scale datasets with relative ease, which has a significant impact on research in the field of 3D pattern analysis, where ad-hoc and diverse datasets are still scarce.

The models are organized into three sets: the query set, the retrieval set, and the training set.





\paragraph{Query set} 
The query set consists of 54 triangle meshes, generated by applying different combinations of relief patterns onto six base surfaces.
Two distinct relief patterns are present on the surface of 40 of these meshes, while the remaining 14 present a single pattern applied across the entire surface.
All models were uniformly sampled with 100,000 vertices.

The six base surfaces were created by sculpting, twisting, and folding a unit square plane in Blender~\cite{blender}.
All base surfaces are topologically equivalent but each of them exhibits exaggerated geometric features such as cavities, folds, and other kinds of occlusions.
The resulting shapes realistically simulate scenarios, where relief patterns are obstructed or not easily isolatable, and a robust texture analysis algorithm must be capable of handling such challenging cases.

\begin{figure}[!ht]
\centering
\includegraphics[width=\linewidth]{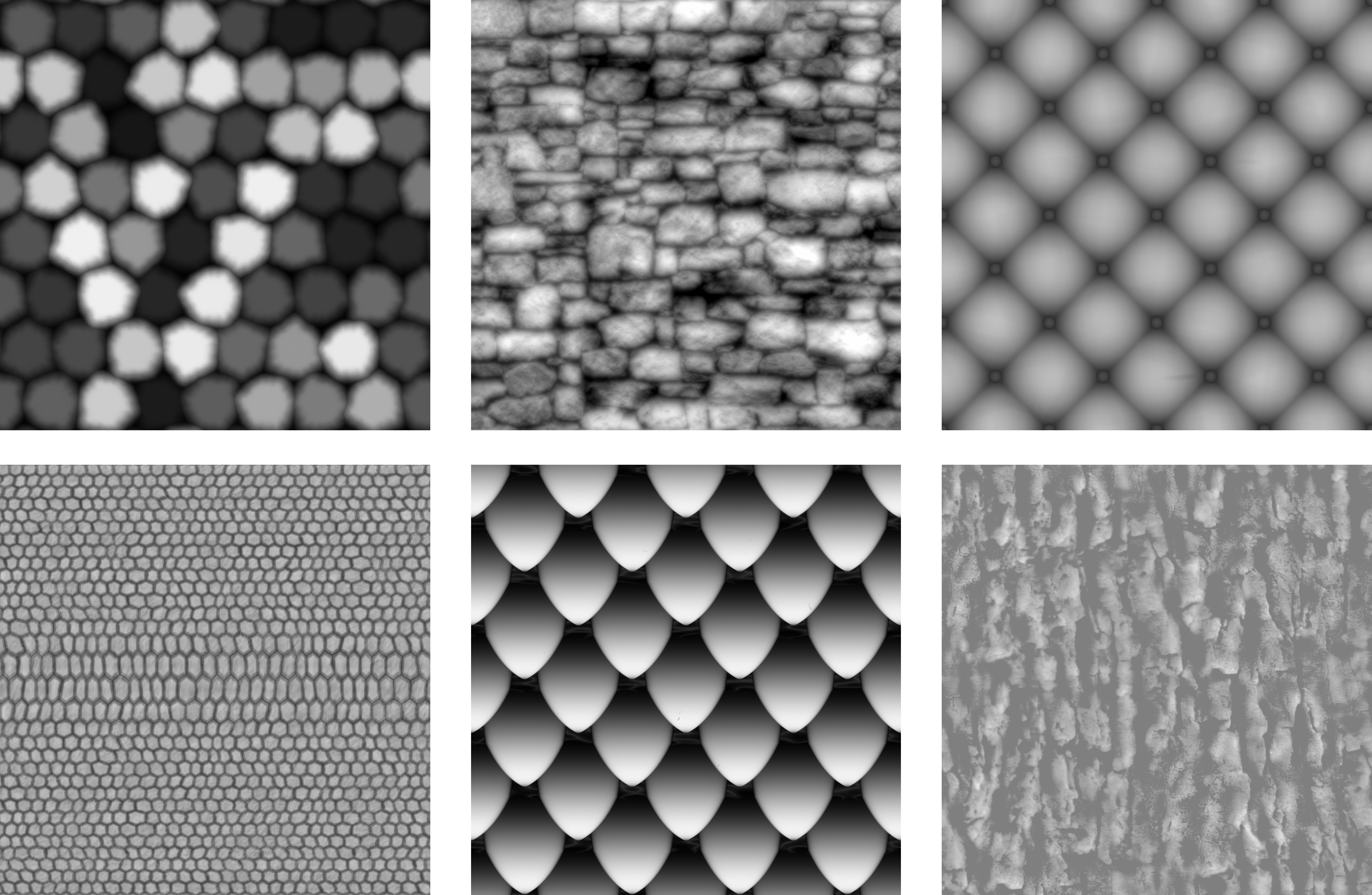}
\caption{Examples of textures selected for generating the query dataset.}
\label{fig:textures}
\end{figure}

The query dataset includes a total of 14 categories of relief patterns, selected from a curated texture collection~\cite{joaopaulopbrtextures}.
These textures were chosen to represent both man-made materials, such as stone walls, padded fabrics, and armor scales, and organic textures, such as snake skin and pine bark.
Some examples of the textures are shown in Figure~\ref{fig:textures}.

These meshes are used as queries to retrieve all models exhibiting similar patterns within the retrieval dataset.

\paragraph{Retrieval set} The retrieval set consists of 300 triangle meshes.
We selected 15 base models from publicly available datasets hosted on platforms such as Polyhaven~\cite{polyhaven} and Sketchfab~\cite{sketchfab}. 
The selected models include both common everyday objects, such as pillows, chairs, and vases, as well as models with more complex geometries, such as the torus, the Stanford dragon, and the Utah teapot.
The diversity of the selected surfaces allows for the evaluation of a method’s capability to analyze relief patterns independently of the underlying geometry, a fundamental property for accurately recognizing local surface details.
To generate the final dataset, various combinations of one, two, or three relief patterns were applied to each base model.
Finally, the resulting meshes were simplified to contain between 100,000 and 200,000 vertices.

\begin{figure*}[!ht]
\centering
\includegraphics[width=0.8\linewidth]{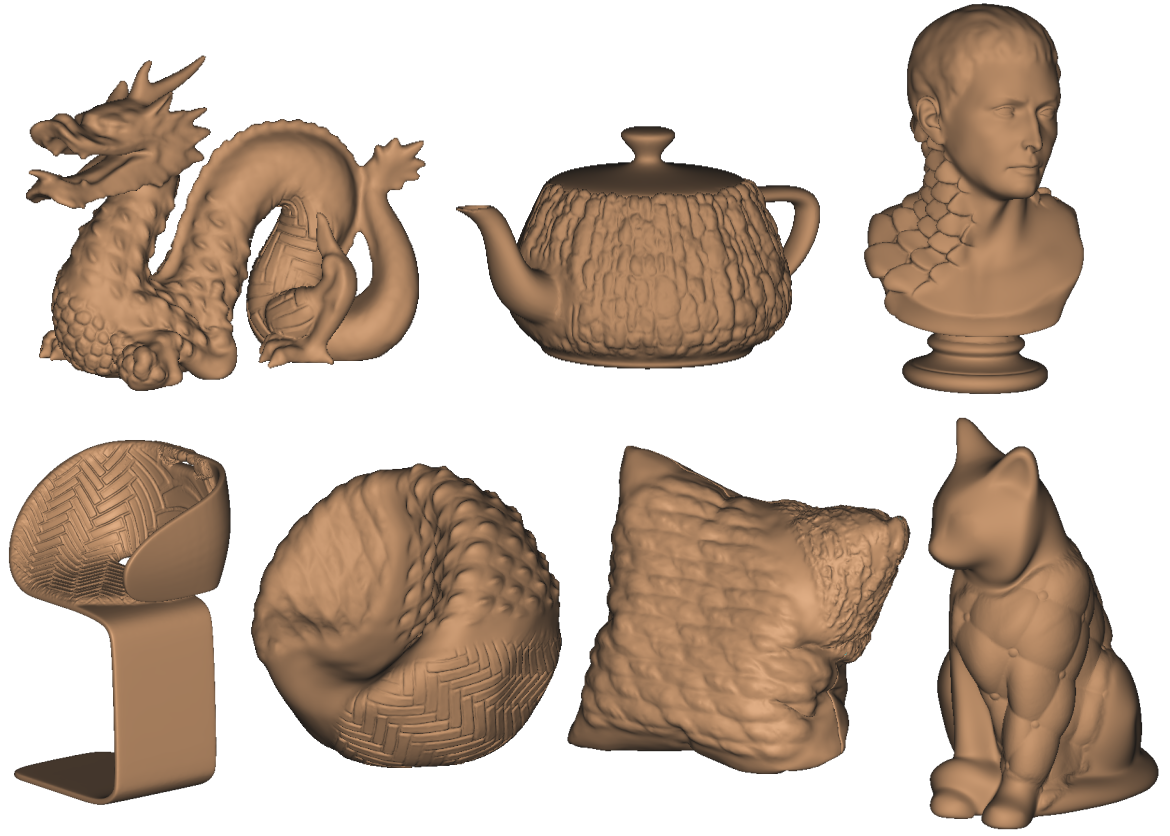}
\caption{Examples of final models from the training and retrieval datasets.}
\label{fig:retrieval-models}
\end{figure*}

\paragraph{Training set} This track includes a training set composed of 700 triangle meshes, generated following the same procedure used for the retrieval set.
What distinguishes the training set is the presence of certain classes of patterns that are not shared with the retrieval set.
This separation was introduced to reflect the need to recognize patterns belonging to previously unseen categories, a situation that frequently arises in real-world applications. 
Some examples of training and retrieval models are shown in Figure~\ref{fig:retrieval-models}.

Ground truth annotation is provided for all models in the training set.
For each mesh, a label is assigned to each face, indicating the relief pattern class associated with it.
The availability of such annotations enables participants to develop segmentation and retrieval methods based on supervised learning.

\section{The participants and the proposed methods}\label{sec:results}
The initial response to the contest was positive, with five groups registering for participation.
By the submission deadline, however, only one group submitted partial results in the form of a membership matrix.
Participants of another group did not complete their submission but conducted preliminary experiments and provided feedback on the challenges encountered during the development of their method.
In the following, we introduce each proposed method along with the corresponding results.

\subsection{KU-3DSeg by Iyyakutti Iyappan Ganapathi, Syed Sadaf Ali, Divya Velayudhan, Maregu Assefa, Naoufel Werghi}\label{sec:iyyakutti}
This method relies on an image classification model to distinguish the relief patterns present in the retrieval meshes.
\begin{enumerate}
    \item \textit{Feature extraction}: each query mesh is processed to extract a local region around a subset of 20,000 randomly selected faces.
    The random sampling of faces is employed to reduce the computational load, since each query model and each retrieval model comprises approximately 100,000 faces and 400,000 faces, respectively.
    For each selected face, neighbouring faces within a dynamically adjusted radius are identified.
    The initial radius is set based on the mean edge length of the mesh faces.
    If the number of neighbours exceeds the specified maximum, the radius is reduced by 25\% in each subsequent attempt until the number of neighbours falls within the desired range.
    This adaptive approach ensures that the number of neighbours is manageable while maintaining the representativeness of the sampled faces~\cite{ganapathi20233dtexsegunsupervisedsegmentation3d, ganapathiunsuperviseddualtransformer, ganapathidetecting3dtexture}.

    The method computes local geometric properties for each selected face and its neighbors, including local depth, face area, centroid, surface variation, normals, and curvature.
    These features are then normalized to ensure consistency and comparability across different faces.
    The resulting descriptors are used to generate a 2D representation of the surface variations that can be fed to the classifier. 
    To do so, the Least Squares Conformal Map (LSCM) algorithm is employed to generate a UV map from the extracted local region.
    The UV map is then rasterized into a grid and the final image is produced by combining different geometric features in a three-channel image.
    
    \item \textit{Classification model training}: 
    in this step, the model learns to distinguish between different types of faces within the query dataset.
    A pre-trained ResNet-50 architecture is fine-tuned on the image features generated in the preceding step, employing the cross-entropy loss and the Adam optimizer.
    The ground truth is implicitly provided by the 14 query meshes that contain a single pattern extended over the entire surface.

    \item \textit{Inference}: for each model in the retrieval dataset, the same set of geometric properties (local depth, area, centroid, surface variation, normals, curvature) is computed for a randomly sampled set of faces.
    The normalized features are fed into the trained classification network to produce a predicted per-face label.

    \item \textit{Label propagation}: each triangle may initially have a label predicted by the classifier, but these labels are often sparse or incomplete (due to the random sampling).
    To improve label coverage and accuracy, a \textit{face adjaceny graph} is constructed for each mesh.
    In this graph, nodes represent mesh faces, and edges connect faces that share at least one vertex.
    This graph structure serves as the foundation for propagating labels across the mesh surface.
    Label propagation is implemented through an iterative, simple weighted voting mechanism~\cite{ganapathigraphclassification}.
    Since we computed 2D features for only a subset of faces, for each unlabeled face, the algorithm calculates influence weights based on three primary geometric criteria: normal similarity, centroid distance, and curvature consistency.
    These weights are combined to determine the most influential neighboring labels, which are then assigned to the unlabeled face if the confidence threshold is met.
    The process continues for a number of iterations or until convergence.
    A fallback mechanism ensures that every face eventually receives a label, even if local consensus is weak.

    \item \textit{Retrieval}: based on the final classification results, relevant 3D models from the retrieval set are retrieved and presented as matches to the query.
\end{enumerate}

\subsection{3D Relief Recognition by Orthographic MultiView Data Characterization and MLP based image classification (OMVMLP) by Ahmed Hazem Youssef, Halim Benhabiles, Adnane Cabani, Ruiwen He, Karim Hammoudi}
This method addresses relief pattern retrieval by characterizing 3D surfaces through multiple orthographic 2D views.
This approach is inspired by the work on pottery retrieval by Benhabiles et al.~\cite{Benhabiles2016}. 
More specifically, six orthographic projections are extracted from each 3D model to ensure comprehensive coverage of the surface.
Each view is then processed through a VGG19 backbone pre-trained on ImageNet~\cite{VGG19}, from which individual feature vectors are extracted.
These vectors are concatenated into a single global descriptor and passed to an MLP classifier that predicts the presence of specific relief patterns from a predefined set. 
Thus, the output of the model is a probability value that the processed mesh shares at least one pattern with the query mesh.

\begin{figure*}[!ht]
\centering
\includegraphics[width=0.9\linewidth]{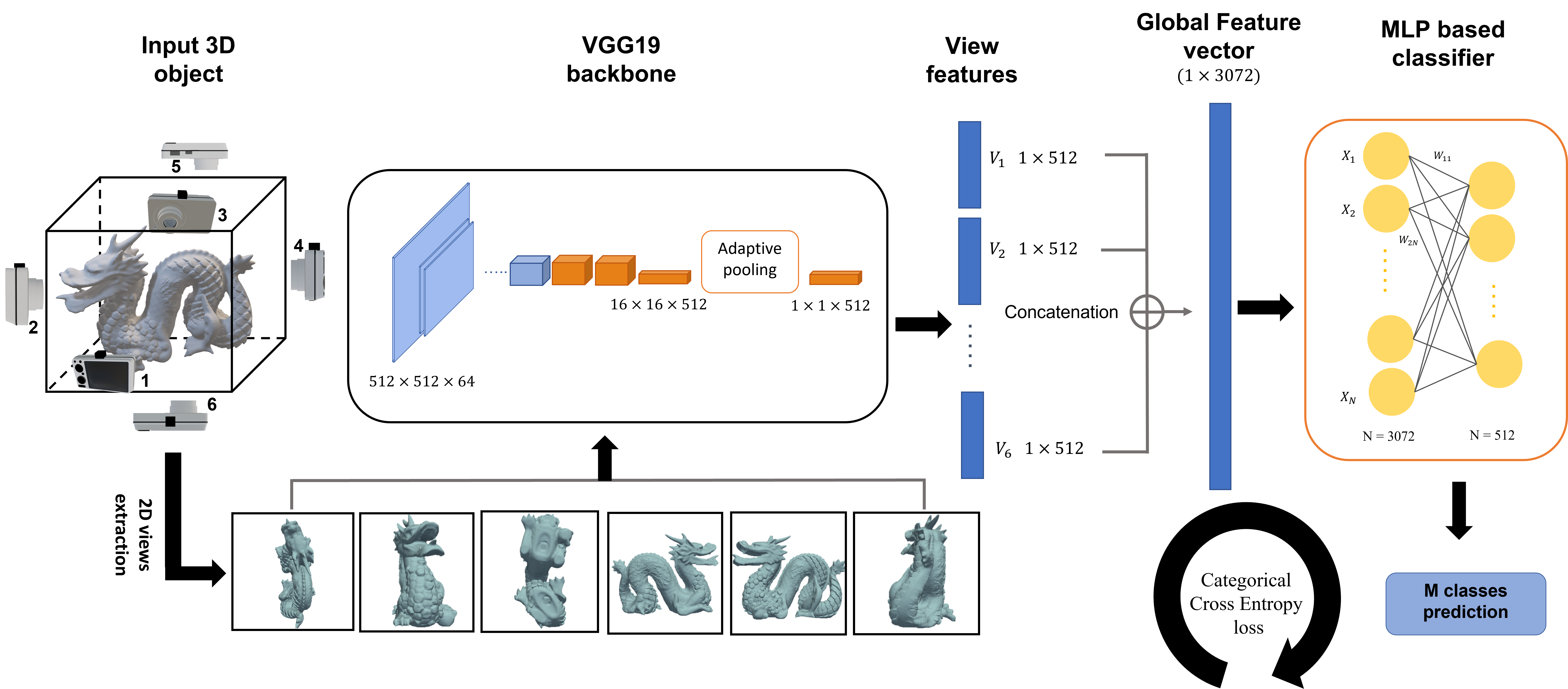}
\caption{Overview of the OMVMLP approach for relief pattern retrieval.}
\label{fig:omvmlp}
\end{figure*}

The whole architecture was trained as a 15-class multi-label classifier corresponding to the relief categories in the training set, using a sigmoid activation in the final layer.
All convolutional layers of VGG19 were frozen except for the last three.
The MLP receives a single 512-dimensional feature vector, obtained by a linear transformation of the concatenated features from all views.
The vector is then processed by a dropout layer, followed by a ReLU activation function and the final output layer.
Training was performed with a batch size of 16 over 70 epochs, using categorical cross-entropy loss and a cosine annealing learning rate scheduler.
After each epoch, performance is evaluated using the F1 score.
The overall network architecture is illustrated in Figure~\ref{fig:omvmlp}

\subsubsection{Membership matrix generation}
Each element of the membership matrix represents the probability that a model in the retrieval set contains at least one relief class present in a given query.
This probability is estimated using the proposed 15-class classifier, which predicts class membership scores for each target mesh.
In addition to the 15 relief categories included in the provided training set, 4 additional relief classes were present in the set of queries. 
As a result, the proposed 15-class classifier is unable to recognize these additional unseen patterns.
To address this limitation, the authors proposed two different strategies:
\begin{itemize}
    \item \textit{Strategy 1}: the 15-class predictor is used to extract a global 512-dimensional feature vector for both query and target meshes.
    For each query-target pair, the similarity between the feature vectors is assessed using the complement of the normalized Euclidean distance.
    The resulting value describes the similarity between two vector representations and is, in turn, interpreted as the probability that the two models exhibit similar relief patterns on their surfaces. 
    
    \item \textit{Strategy 2}: the classifier is adapted by replacing its original 15-class output layer with a 4-class output layer, which is then fine-tuned on the set of queries.
    This adapted classifier is then used to estimate the probability that a target model contains one of the four additional relief classes.
\end{itemize}

\subsubsection{Results and discussions}
The authors of OMVMLP submitted two membership matrices, one for each of the two strategies described above.
The reported results therefore reflect only the ability to retrieve meshes with patterns similar to those in the query, without providing any localization of the pattern on the surface.
The evaluation metrics selected are a combination of measures previously adopted in earlier SHREC tracks~\cite{SHREC17,MOSCOSOTHOMPSON2020199,10.5555/3290638.3290649}: Nearest Neighbor (NN), First tier (FT), Second tier (ST), normalized Discounted Cumulated Gain (nDCG), mean Average Precision (mAP), e-Measure (e), Receiver Operating Characteristic (ROC) curve, and AUC value. 
Before computing the evaluation metrics, we removed the results associated with queries for which no relevant mesh exists in the retrieval dataset.
This allowed for interpretable measurements under the standard working conditions of the proposed algorithm, i.e., when the query shares relief classes with models in the retrieval set. 

\begin{figure}[!ht]
\centering
\includegraphics[width=0.8\linewidth]{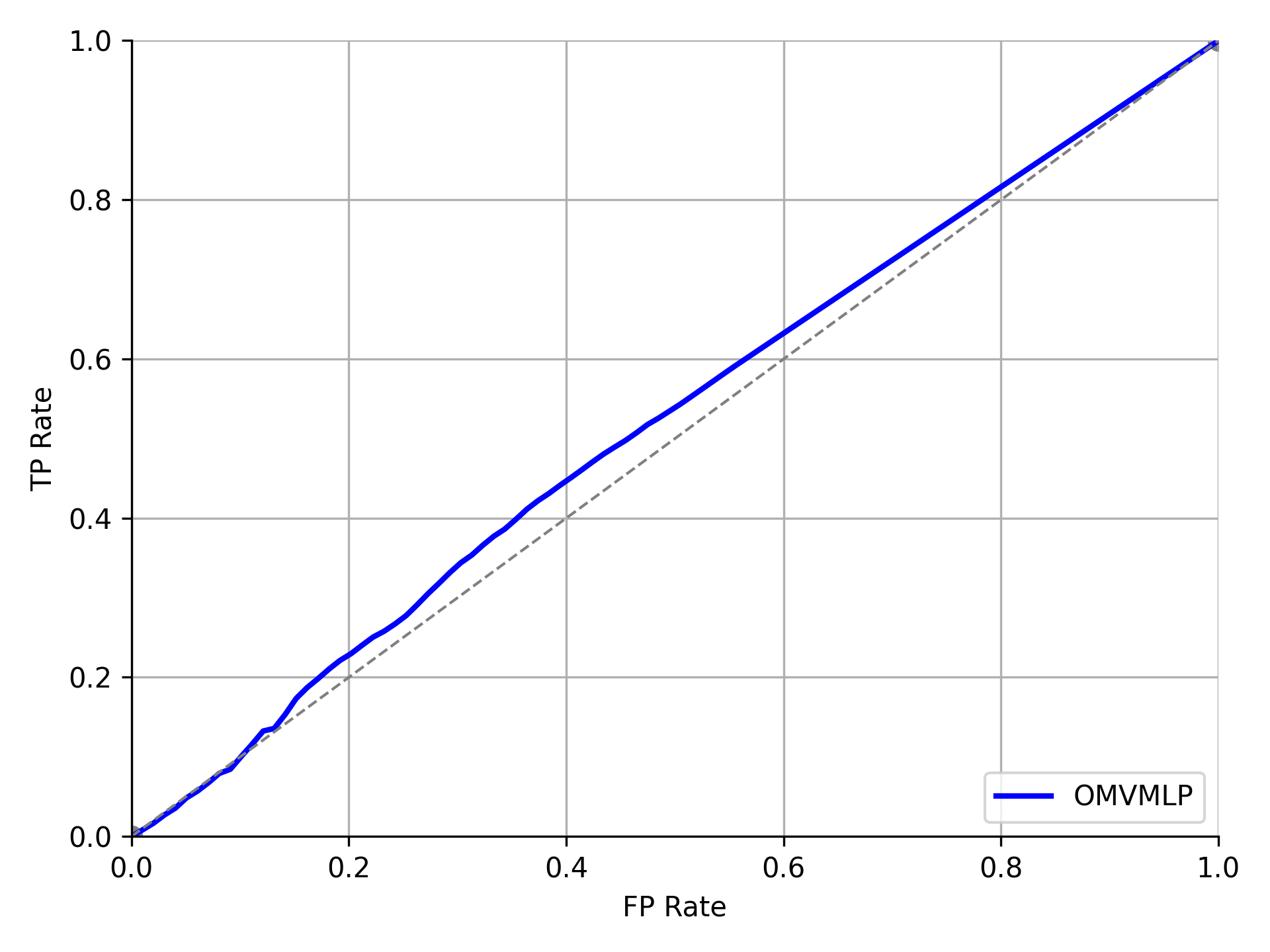}
\caption{The ROC curve for both OMVMLP runs. The dashed diagonal represents the baseline performance of a random classifier. As described in Fig.~\ref{tab:omvmlp-results}, a single ROC curve is sufficient to evaluate both runs.}
\label{fig:omvmlp-results}
\end{figure}

In Table~\ref{tab:omvmlp-results} we report the performance of the OMVMLP technique, evaluated using the aforementioned measures.
Considering the exclusion of inferences from queries with no corresponding retrieval meshes, the reported metrics reflect the performance across both runs (strategy 1 and 2).
The ROC curve is shown in Figure~\ref{fig:omvmlp-results}.
Overall, the method demonstrated unsatisfactory retrieval performance.
The AUC value, equal to 0.523, indicates performance only slightly better than that of a random classifier.
For comparison, similar values were reported for the techniques \textit{Augmented point pair feature descriptor aggregation with fisher kernel} (APPFD-FK) and \textit{Signature quadratic form distance and PointNet} (PointNet+SQFD) in SHREC 2020~\cite{MOSCOSOTHOMPSON2020199}.


To justify the results obtained by OMVMLP, it is important to note that the method analyzes six orthographic projections of the entire 3D model.
As a consequence, the VGG19 network primarily extracts features related to the object as a whole.
In the frequent scenario where geometric textures are not fully visible from the six predefined viewpoints, or are entirely hidden (e.g., patterns located inside vases), the extracted features are unlikely to carry information relevant to the retrieval task.

Taking inspiration from the DFE method~\cite{MOSCOSOTHOMPSON2020199}, which shares OMVMLP's general approach, it may be beneficial to aggregate features from intermediate layers of the network.
Intermediate representations tend to capture more localized and spatially descriptive information, which is more suitable for representing the projected relief patterns.
In contrast, relying solely on the final output vector of the network may result in overly compact representations that fail to discriminate fine-grained geometric details.
Equally important is the heuristic used for selecting the projection viewpoints.
In cases where the surface contains multiple relief pattern classes, the viewpoints must be chosen so as to effectively capture all types of textures.
It is evident that, in the absence of any prior knowledge about the number and spatial extent of such textures, this step becomes a major bottleneck for image-based analysis algorithms.

\begin{table}[h]
    \centering
    \small
    \begin{tabular}{|c|c|c|c|c|c|c|}
    \hline
    NN & FT & ST & mAP & nDCG & e & AUC \\ \hline
    0.313 & 0.219 & 0.430 & 0.218 & 0.655 & 0.364 & 0.523 \\
    \hline
    \end{tabular}
    \caption{Nearest Neighborhood (NN), First Tier (FT), Second Tier (ST), mean Average Precision (mAP), normalized Discounted Cumulated Gain (nDCG), e-measure (e) and Area Under the Curve (AUC) values of the OMVMLP runs. Values goes from 0, to 1. The higher the value is, the better the method's performance. The values in the table were computed by excluding the results of queries without any relevant model in the retrieval dataset. Since the two strategies differ only in their performance on the aforementioned query classes, the values reported in the table reflect the outcome of both runs.}
    \label{tab:omvmlp-results}
\end{table}

\section{Relief pattern analysis challenges}\label{sec:analysis-challenges}
Relief pattern recognition has attracted significant interest from the research community, as evidenced by the impressive performance and diversity of solutions proposed in the SHREC 2020 track on the topic~\cite{MOSCOSOTHOMPSON2020199}.
As highlighted in the track report, most methods pre-process the surface by extracting one or more representative patches.
This operation serves a dual purpose: to reduce the computational burden of computing descriptors over the entire surface, and to isolate a representative region of the relief pattern that is as planar as possible.
Patch planarity is important to decouple the relief pattern from global geometric variations.
Patch extraction can be performed in various ways, but typically involves the following steps:

\begin{enumerate}
    \item Normalization of the model's scale and alignment along the canonical axes for the ease of rendering.
    \item Projection onto an image plane of the textured surface region that maximizes area coverage, while satisfying predefined planarity conditions (see the DFE and DPML methods~\cite{MOSCOSOTHOMPSON2020199}).
\end{enumerate}   

The extraction of planar patches becomes more challenging on surfaces that contain numerous high-curvature regions, such as small handles, thin parts, or fine local details.
In order to isolate a patch that contains the relief pattern, it may be necessary to reduce the patch size, at the cost of a lower signal-to-noise ratio.

In general, the existing approaches cannot be easily adapted to solve the task of relief pattern localization.
This is because retrieval techniques typically aggregate information over a representative patch to identify the type of pattern present.
In contrast, surface segmentation requires predicting the pattern class at the level of individual faces.
Moreover, the heuristics used to project the patch onto an image are not inherently capable of detecting the presence of multiple pattern classes.
The datasets released with SHREC 2017~\cite{SHREC17} and SHREC 2020~\cite{MOSCOSOTHOMPSON2020199} contain a single type of pattern extended across the entire surface, making the use of such techniques appropriate.
In contrast, the datasets released with SHREC 2018~\cite{SHREC18} and the present track feature multiple pattern classes on the same surface, rendering this approach no longer applicable.
Further confirmation of this is provided by the performance of the OMVMLP method.
We believe, however, that its results could be significantly improved by employing a more guided search for the most informative view of the model to render and analyze. 

\begin{figure*}[!ht]
    \centering  
    \begin{subfigure}[b]{0.23\textwidth}
        \includegraphics[width=\textwidth]{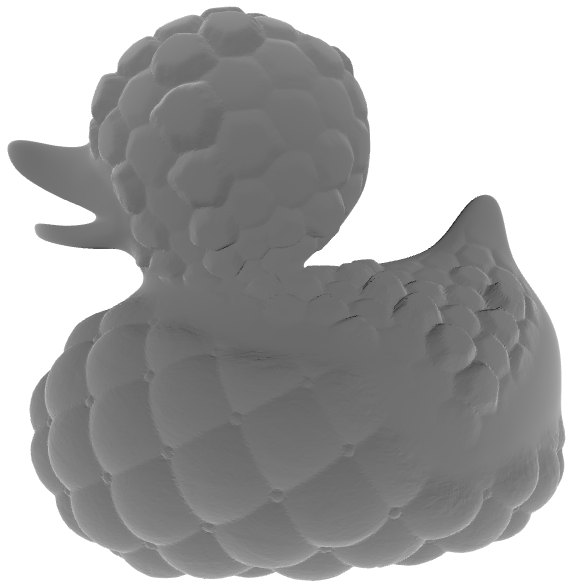}
        \caption*{Original 1}
    \end{subfigure}
    \begin{subfigure}[b]{0.23\textwidth}
        \includegraphics[width=\textwidth]{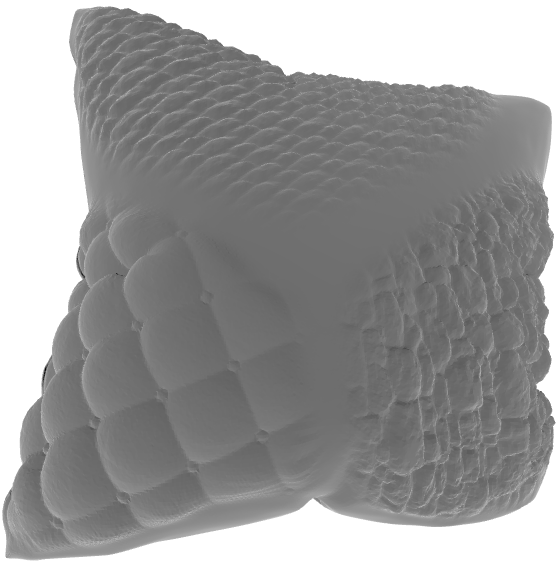}
        \caption*{Original 2}
    \end{subfigure}
    \begin{subfigure}[b]{0.23\textwidth}
        \includegraphics[width=\textwidth]{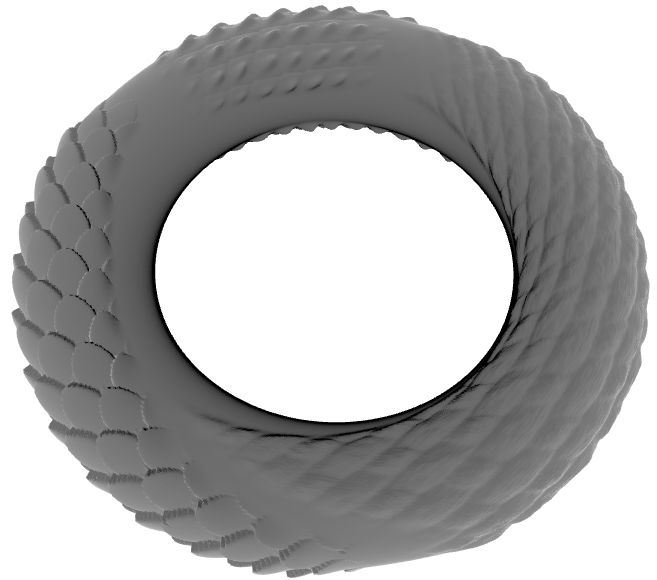}
        \caption*{Original 3}
    \end{subfigure}
    \begin{subfigure}[b]{0.23\textwidth}
        \includegraphics[width=\textwidth]{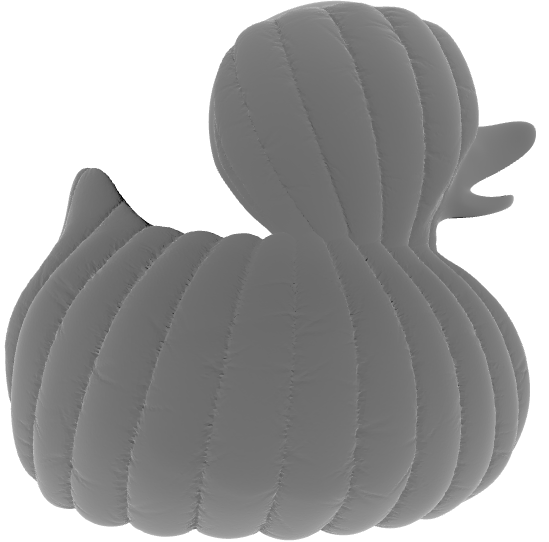}
        \caption*{Original 4}
    \end{subfigure}

    \vspace{1em}

    \begin{subfigure}[b]{0.23\textwidth}
        \includegraphics[width=\textwidth]{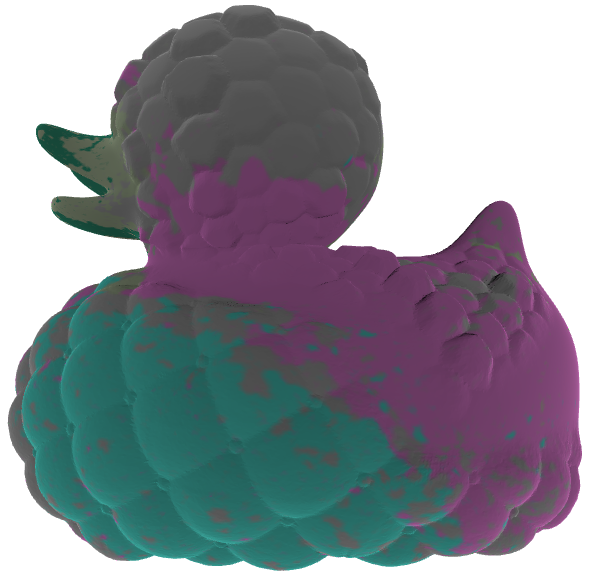}
        \caption*{Segmented 1}
    \end{subfigure}
    \begin{subfigure}[b]{0.23\textwidth}
        \includegraphics[width=\textwidth]{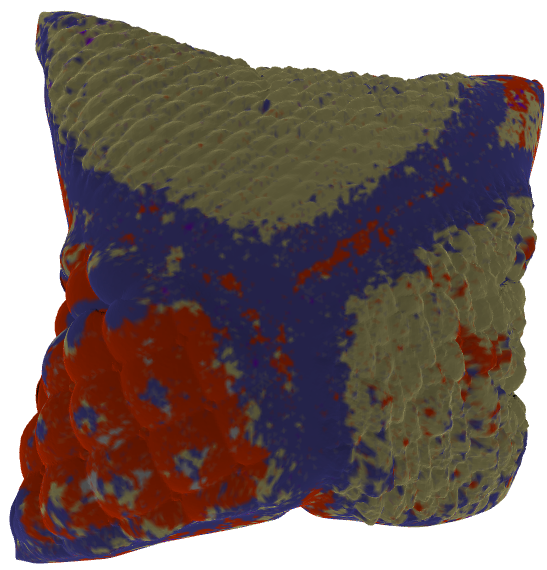}
        \caption*{Segmented 2}
    \end{subfigure}
    \begin{subfigure}[b]{0.23\textwidth}
        \includegraphics[width=\textwidth]{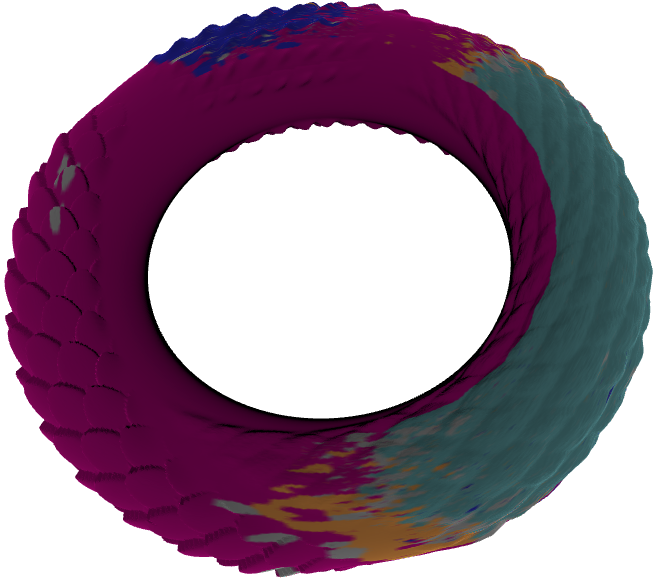}
        \caption*{Segmented 3}
    \end{subfigure}
    \begin{subfigure}[b]{0.23\textwidth}
        \includegraphics[width=\textwidth]{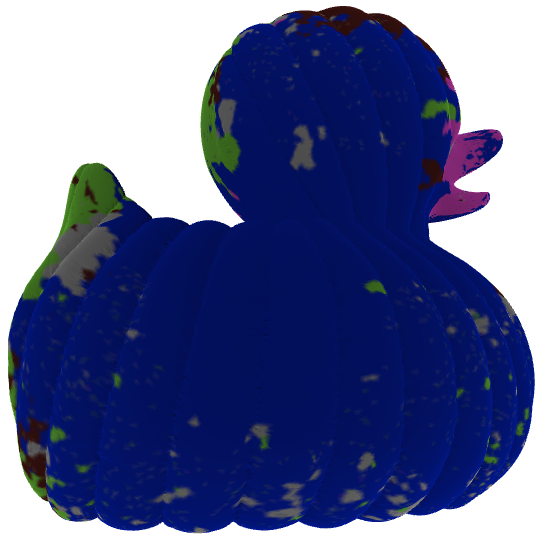}
        \caption*{Segmented 4}
    \end{subfigure}
\caption{Top row: original models. Bottom row: corresponding segmented outputs from KU-3DSeg. This result is provided by the group of Iyyakutti et al.}
\label{fig:qualitative-results}
\end{figure*}

\begin{figure*}[!ht]
    \centering
    \begin{subfigure}[b]{0.2\textwidth}
        \includegraphics[width=\textwidth]{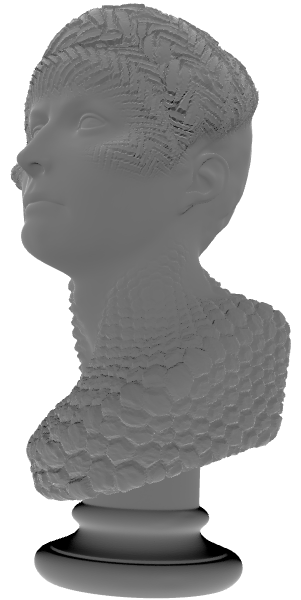}
    \end{subfigure}
    \begin{subfigure}[b]{0.2\textwidth}
        \includegraphics[width=\textwidth]{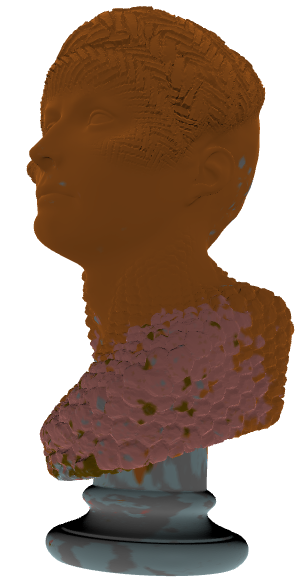}
    \end{subfigure} 
    \begin{subfigure}[b]{0.3\textwidth} 
        \includegraphics[width=\textwidth]{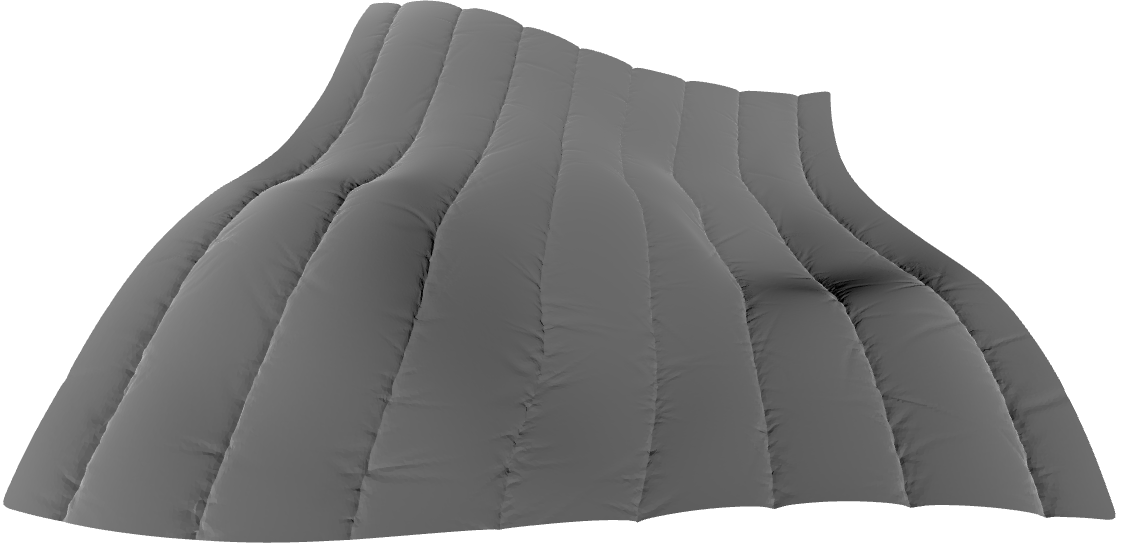}
    \end{subfigure} 
    \caption{On the left, an example of a retrieval model is shown. In the center, the result of the pattern retrieval is displayed, with the different regions segmented. The lower section of the bust does not correspond to any relief pattern; however, its shape resembles Query 35 on the right due to the radius utilised to extract 2D features. This result is provided by the group of Iyyakutti et al.}
    \label{fig:pattern-size}
\end{figure*}

Below, we summarize the main challenges reported by the participants.
\begin{itemize}
    \item \textit{Memory and computational complexity}: 
    the high resolution of the models in the dataset represented one of the main challenges in the development of each method.
    The authors of OMVMLP were unable to solve either of the two proposed tasks and circumvented the issue of per-face annotation by designing an architecture for retrieving models with relevant relief patterns.
    As a result, model resolution does not impact the computational complexity of their approach, as it relies entirely on generating 2D renderings of the 3D models.

    The authors of KU-3DSeg developed an architecture capable of identifying whether and where geometric textures are present.
    However, the feature extraction step involves generating a grid for each face of the mesh.
    This operation is computationally dependent on both the number of faces in the mesh and the size of the neighborhood from which features are extracted.
    To make the problem computationally tractable, only a randomly sampled subset of mesh faces is processed. 

    \item \textit{Interface between patterns}: 
    the label propagation mechanism described in Section~\ref{sec:iyyakutti} provides a mean of compensating for the random downsampling of mesh faces described previously.
    In Figure~\ref{fig:qualitative-results}, we observe that the segmented regions tend to group similar labels reasonably well.
    However, challenges arise, particularly near the boundaries between different patterns.
    The computation of per-face 2D features takes into account a neighborhood of adjacent faces, whose size is set based on the mean edge length of the mesh.
    As a result, faces located at the interface between two relief patterns often include neighboring faces belonging to both textures.
    This overlap leads to ambiguity in the final face classification.
    An example of this issue is shown in Figure~\ref{fig:qualitative-results}, first column.
    Here, the region marked in pink lies between two relief patterns, colored gray and cyan.
    Ideally, the pink label should correspond to the plain surface separating the two patterns.
    However, due to the inclusion of mixed-neighborhood features at the boundary, the pink label has propagated into areas belonging to both the gray and cyan regions.
    This highlights a limitation of the current feature design in effectively handling pattern boundaries.

    \item \textit{Local-global nature of relief patterns}: 
    a known issue, already observed in previous works, concerns the handling of relief patterns with significantly different spatial scales~\cite{PAOLINI2024104020}.
    A pattern may, for example, be characterized by fine details along one direction, while remaining smooth (i.e., lacking repeated corrugations) along another.
    When analyzing patterns with large spatial extent, increasing the neighborhood radius may lead to misclassification in which decorative elements or structural parts of the model are mistaken for textured regions.
    This issue is illustrated in Figure~\ref{fig:pattern-size}, where the base of the bust model shown in the first two columns is erroneously identified as similar to Query 35 (third column).
    This misclassification occurs because the extended radius includes features that resemble those of the query, even in areas that are otherwise plain or lack any geometric texture.
    This variability in the spatial characteristics of relief patterns calls for both local and global descritptions of the object, as well as a criterion capable of reliably distinguishing repeated features from decorative or structural ones.
\end{itemize}

These issues do not reflect inherent difficulties in geometric texture recognition itself, but rather expose the limitations of current mesh analysis methods.
What emerges is the need for developing fundamentally different approaches from those proposed so far, which still remain largely based on texture image analysis techniques.

\section{Challenges of the dataset}\label{sec:dataset-challenge}
The models presented in this challenge, although free from noise or artifacts typically produced by laser scan acquisition, exhibit a set of characteristics that make the contest particularly challenging.
\begin{itemize}
    \item \textit{Multiple 3D textures}. 
    The choice of applying more than one geometric texture to each surface reflects the most general case, in which both the number and characteristics of the relief patterns are unknown a priori.
    With the development of increasingly large-scale 3D databases~\cite{objaverseXL}, it becomes necessary to implement efficient 3D-to-3D data query systems that enable the retrieval of models based on their surface geometric features.
    
    \item \textit{Complex geometries}. 
    The base models used to construct the training and retrieval sets were selected with the goal of providing a wide variety of geometric configurations.
    They range from models with common shapes (e.g., human figures), to more primitive forms (e.g., a soccer ball), up to highly detailed meshes with self-occlusions, such as the Stanford dragon.
    We expect that simple or relatively planar models can be more easily processed by projection-based methods. 
    In contrast, closed surfaces such as the Utah teapot, which exhibit geometric ``thickness,'' may conceal relief patterns within and make it difficult, if not impossible, to effectively apply projection-based techniques.
    The inclusion of complex geometries is thus intended to encourage the development of methods that extract information directly from the surface~\cite{ganapathi20233dtexsegunsupervisedsegmentation3d,PAOLINI2024104020,6954469}.
    
    \item \textit{Texture selection}. 
    The set of texture images used to generate the relief patterns includes regular spatial variations (e.g., floors and fabrics), quasi-regular textures (e.g., animal skins and stone walls), and irregular ones (e.g., tree bark and metallic materials).
    Such imperfections can be treated as a source of noise, although they cannot, in principle, be modeled in the same way as noise introduced by 3D acquisition devices.
    The bark texture serves as an edge case, as it lacks any spatial periodicity.
    When dealing with this type of data, a multi-resolution analysis approach, such as the Dual Tree Complex Wavelet Transform (DT-CWT)~\cite{OTHMANI20132144}, may prove beneficial.
    
    \item \textit{Disjoint set of patterns}. 
    Some texture classes are exclusive to each of the three released datasets.
    This design choice aims to evaluate the generalization capabilities of methods when faced with previously unseen patterns.
    In general, learning-based models tend to struggle more than feature-based methods when recognizing data not encountered during training.
    This limitation may arise from several factors, including network overfitting, the use of pre-trained models not well suited to the task, or sub-optimal parameter configurations (e.g., the size of the extracted patch).
\end{itemize}

\section{Possible research directions}\label{sec:directions}
The methods reported here reflect the two main strategies for 3D pattern analysis: transfer learning approaches based on pre-trained, image-based neural networks, and feature-based methods.
Both strategies have achieved strong performance in classification and retrieval tasks involving similar geometric reliefs~\cite{SHREC17,10.1007/978-3-642-15561-1_11}, although there has been a growing shift towards machine learning techniques specifically designed or adapted for these tasks.

In contrast, the outcome of this track, combined with the challenges reported by the participants, suggests several research directions that deserve increased attention.
\begin{itemize}
    \item \textit{Development of task-specific 3D methods}. 
    With the success of deep learning algorithms in image recognition, many authors have experimented with adapting or extending 2D approaches to 3D domains. 
    However, non-Euclidean data (such as meshes, point clouds, or graphs) lack the regular grid structure of images, making operations like convolution and pooling non-trivial to extend to 3D. 
    Over the last decade, various definitions of convolution and pooling mechanisms for non-Euclidean data have emerged~\cite{9674191}, with applications to tasks such as functional correspondence, molecular and human body segmentation, and shape classification~\cite{sharp2022diffusionnetdiscretizationagnosticlearning}. 
    Methods developed in the context of geometric deep learning~\cite{Bronstein2017} may provide a foundation for more robust algorithms capable of handling noise, self-occlusions (without relying on 2D projections), and generalizing to unseen data.

    In the domain of traditional geometric processing, there are also several algorithms for extracting quasi-regular surface structures that could be explored and adapted to this problem.
    For example, under broad conditions, the optimization framework proposed in~\cite{Huang2014} is capable of detecting approximately regular geometric elements distributed across surface regions.
    
    \item \textit{Foundation models}. 
    Powerful backbones such as ResNet~\cite{7780459}, DenseNet~\cite{8721151}, VGG~\cite{VGG19}, and EfficientNet~\cite{tan2020efficientnetrethinkingmodelscaling} have enabled many 3D downstream tasks to be formulated as latent feature analysis problems, reducing the need for expensive training.
    More recently, the emergence of visual foundation models (e.g., CLIP~\cite{radford2021learningtransferablevisualmodels}, DINOv2~\cite{oquab2024dinov2learningrobustvisual}, iBOT~\cite{zhou2022ibotimagebertpretraining}) has shown impressive generalization capabilities across a wide range of applications.
    A recent study~\cite{el2024probing} conducted a series of experiments to assess the 3D awareness of these models.
    The authors conclude that while most visual foundation models learn representations that encode depth and visible surface orientation, they struggle when matching across significantly different viewpoints, an issue closely related to the Janus problem observed in diffusion-based 3D reconstruction methods.
    Despite this, the application of visual foundation models to relief pattern recognition remains largely unexplored.
    A natural extension would involve developing 3D backbones aligned with multimodal features, such as image-text pairs derived from visual-language models~\cite{xue2024ulip2scalablemultimodalpretraining, liu2023openshape}.
    The impressive results achieved in tasks like zero-shot 3D shape classification and multimodal 3D shape retrieval suggest that similar capabilities could extend to geometric texture analysis, although adapting such models to support segmentation remains a non-trivial challenge.
    
\end{itemize}

\section{Conclusions and future insights}\label{sec:conclusions}
The objective of this track was twofold: to determine whether and where geometric textures are present in synthetic triangular meshes, and to segment surfaces containing two distinct relief patterns.
Unfortunately, no participants submitted complete solutions for either task.
One group provided valuable feedback based on preliminary testing on the dataset, while another submitted results from their method for retrieving meshes with relief patterns similar to those in the query.
Although the latter results were unsatisfactory, they highlight an inherent difficulty in applying image-based methods when the complexity of the models increases.
High mesh resolution, the challenge of defining the extent of each pattern, and the difficulty of distinguishing certain texture classes from background geometry or noise emerged as the most common issues encountered by participants.

We proposed several potential research directions.
Direct approaches based on handcrafted descriptors offer the advantage of not requiring large amounts of training data and allow for greater transparency in their behavior.
However, the versatility of these algorithms depends on a deep understanding of the problem, the types of patterns involved, and the characteristics of the 3D surfaces under analysis.
On the other hand, leveraging the capabilities of recent foundation models may represent a promising path toward developing techniques capable of recognizing a wide variety of texture classes.
The main limitation in using foundation models lies in the ability of the pre-trained 3D backbone to extract and disentangle features across multiple spatial scales, thereby enabling fine-grained analysis of the 3D object.
For both categories of techniques, the high-resolution of 3D models remain a critical challenge that must be addressed in order to apply pattern recognition algorithms to increasingly complex and realistic surfaces. 

Finally, this work introduces a new benchmark for the development and evaluation of future algorithms.
The released training data (700 models) represent one of the largest datasets specifically created for the task of relief pattern analysis.


\printbibliography
\end{document}